\documentstyle{mn}
\input epsf

\begin{document}

\journal{Preprint astro-ph}

\title{\bf A possible contribution to CMB anisotropies at high $\ell$
from primordial voids}

\author[Louise~M.~Griffiths, Martin~Kunz and Joseph~Silk]{Louise M. Griffiths$^{1,2}$, Martin Kunz$^1$ and Joseph Silk$^1$\\
$^1$Astrophysics, University of Oxford, Denys Wilkinson Building, Keble Road, Oxford OX1 3RH, England.\\
$^2$Astrophysics, UNSW, Sydney, NSW 2052, Australia.}

\maketitle


\begin{abstract}
We present preliminary results of an analysis into the effects of
primordial voids on the cosmic microwave background (CMB). We show
that an inflationary bubble model of void formation predicts excess
power in the CMB angular power spectrum that peaks between $2000 <
\ell < 3000$.  Therefore, voids that exist on or close to the last
scattering surface at the epoch of decoupling can contribute
significantly to the apparent rise in power on these scales recently
detected by the Cosmic Background Imager (CBI).
\end{abstract}


\section{Introduction}
One of the primary goals of modern cosmology is to gain an
understanding of the formation and evolution of structure in the
universe.  Analyses of redshift surveys such as the 2-degree field
galaxy redshift survey (2dFGRS, Peacock et al. 2001) suggest that
there are large volumes of relatively empty space, or $voids$, in the
distribution of galaxies. It seems that the universe is made up of a
network of voids with most galaxies tending to be found in two
dimensional sheets or filaments that surround these under-dense
regions.

In the hierarchical model of structure formation, gravitational
clustering is responsible for emptying voids of mass and galaxies
(Peebles 1989). Simulations of the standard cold dark matter (CDM)
model predict significant clumps of matter within voids that are
capable of developing into observable bound objects (Dekel \& Silk
1986; Hoffman et al. 1992).  Peebles gives an in--depth discussion of
the contradictions of this prediction with observation (Peebles 2001).
He argues that the inability of the CDM models to produce the observed
voids constitutes a true crisis for these models. Additionally,
recently announced deep field observations from the Cosmic Background
Imager (CBI) (Mason et al. 2002) show excess power on small angular
scales, $\ell > 2000$, in the cosmic microwave background (CMB).

It may be possible to explain the observations by postulating the
presence of a void network originating from primordial bubbles of true
vacuum that nucleated during inflation (La 1991; Liddle \& Wands 1991;
Turner et al. 1992; Occhionero \& Amendola 1994; Amendola et
al. 1996).  In this scenario, the first bubbles to nucleate are
stretched by the remaining inflation to cosmological scales. The
largest voids may have had insufficient time to thermalize before
decoupling and may persist to the present day.  If voids exist at
recombination they will leave an imprint on the cosmic microwave
background. On the other hand, if they formed much later, their effect
on the CMB will be negligible and will not be observed with the
current generation of experiments.

The effects from primordial voids on the CMB have been investigated by
a number of authors (Thompson \& Vishniac 1987; Sato 1985;
Mart\'{\i}nez-Gonz\'alez et al. 1990; Mart\'{\i}nez-Gonz\'alez \& Sanz
1990; Liddle \& Wands 1992; Panek 1992; Arnau et al. 1993;
M\'esz\'aros 1994; Fullana et al. 1996; M\'esz\'aros \& Moln\'ar 1996;
Shi et al. 1996; Baccigalupi et al. 1997; Amendola et al. 1998;
Baccigalupi et al. 1998).  The most complete investigation was carried
out by Sakai et al. (1999) who modeled the effect for a distribution
of equally sized voids.  

In this Paper, we use a simple inflationary bubble model to show that
if the voids that we see in galaxy surveys today existed at the epoch
of decoupling, they would contribute significant additional power to
the CMB angular power spectrum between $2000 < \ell < 3000$.  Unlike
previous analyses, we develop a general method that allows the
creation of maps and enables us to consider an arbitrary distribution
of void sizes.  We model a power law distribution of void sizes as
predicted by inflation (La 1991) and also take into account the finite
thickness of the last scattering surface which suppresses part of the
contribution from small voids.

\section{Void networks in theory and observation}

\subsection{Predictions of the inflationary bubble model}
In the extended inflationary model (La \& Steinhardt 1989; see Kolb
1991 for a review), true vacuum bubbles nucleate during inflation in
first order phase-transitions.  This model predicts a distribution of
bubble sizes greater than a given radius $r$ of the form,
\begin{equation}
N_B(>r) \propto r^{-\alpha} . \label{eqvoidsize}
\end{equation}

Typically, extended inflation is implemented within the framework of a
Jordan-Brans-Dicke theory (Brans \& Dicke 1961). In this case, the
exponent $\alpha$ is directly related to the gravitational coupling
$\omega$ of the scalar field that drives inflation,
\begin{equation}
\alpha = 3 + \frac{4}{\omega + 1/2} .
\end{equation}
Values of $\omega>3500$ are required by solar system experiments (Will
2001), although models have been proposed that either suppress or hide
the present value of $\omega$ (La et al. 1989, Holman et al. 1990).
The main driving force behind these models is that a low $\alpha$ can
lead to large effects on the CMB if arbitrarily large voids are
allowed (see Liddle \& Wands 1991 for a review).  The normalization of
the bubble size distribution also depends on $\omega$ as well as on
the energy scale of inflation.

Once formed, the bubbles will expand and form a shock wave on their
boundary with the surrounding matter. After inflation ends, matter
will start to flow relativistically back into the freshly created
underdensities.  However, cold dark matter only travels minimally into
the void, since it becomes non-relativistic early on (Liddle \& Wands
1992).  Gravitational collapse of CDM will begin as normal at
equality, further emptying any persisting voids. We expect baryonic
matter to be pushed much further into the void as it is tightly
coupled to relativistic photons until the epoch of decoupling when it
will begin to gravitationally collapse back onto the CDM.


\subsection{Void detections in redshift surveys}

A number of different void finder algorithms have been developed to
detect voids in redshift surveys (Kauffman \& Fairall 1991; Kauffman
\& Melott 1992; Ryden 1995; Ryden \& Melott 1996; El-Ad \& Piran 1997;
Aikio \& M\"ah\"onen 1998).  So far, such algorithms have been used to
search for voids in the first slice of the Center for Astrophysics
(CfA) redshift survey (Slezak et al. 1993), the Southern Sky Redshift
Survey (SSRS) (Pellegrini et al. 1989, El-Ad et al. 1996), the
Infra-Red Astronomical Satellite (IRAS) 1.2 Jy survey (El-Ad et
al. 1997), the Las Campanas Redshift Survey (LCRS) (M\"uller et
al. 2000), the Updated Zwicky Catalogue (UZC) (Hoyle \& Vogeley 2002)
and the Point Source Catalogue redshift (PSCz) survey (Plionis \&
Basilakos 2001, Hoyle \& Vogeley 2002). These investigations indicate
that 30-50\% of the fractional volume of the universe is in the form
of voids of underdensity $\delta \rho/\rho < -0.9$, in line with the
inflationary model predictions. These voids range in radius from
$r_{\rm min}$ = 10 $h^{-1}$ Mpc to $r_{\rm max}$ = 20-30 $h^{-1}$ Mpc.


\section{The phenomenological void network model}

We model the voids seen today as spherical underdensities of $\delta
\rho/\rho = -1$.  Each void is bounded by a thin wall containing the
matter that is swept up during the void expansion.  This forms a
compensated void.  We take the background universe to be an
Einstein--de Sitter (EdS) cosmology, which is a good approximation
since the majority of the effect on the CMB comes from voids on or
close to the last scattering surface (LSS) and the universe tends
towards an EdS cosmology at early times. Maeda \& Sato (1983) and
Bertschinger (1985) use conservation of momentum and energy
respectively to show that these compensated voids will increase in
radius $r_v$ between the onset of the gravitational collapse of matter
at equality and the present day such that,
\begin{equation}
\label{e:vexpand}
r_v(\eta) \propto \eta^{\beta} \,,
\end{equation}
with $\beta \approx 0.39$ and where $\eta$ is conformal time.

In this Paper, we consider a phenomenological primordial void model
that is based on the predictions of extended inflation with parameters
chosen to be in agreement with current redshift survey observations;
a full analysis of a larger family of models will be presented
in a subsequent paper.
Motivated by the inflationary scenario, we assume
a power-law distribution of void sizes in the universe today, 
as given in equation (\ref{eqvoidsize}). 
We further assume that the mechanism creating the voids imposes an upper
cut-off on the size distribution. A possible mechanism for this cut-off
could be that the tunneling probability of inflationary bubbles
is modulated through the coupling to another field. 
We can therefore go to
the limit of large $\omega$, leading to a spectrum of void sizes with
$\alpha = 3$. 
Apart from avoiding problems
with well--established local measurements of gravity, this assumption
allows us to match the observed upper limit on void sizes from the
galaxy redshift surveys.

The minimal present void size is chosen to agree with
redshift surveys, $r_{\rm min}=10\, h^{-1}$ Mpc.  
 For the maximal radius, we choose the average value 
that is found,
$r_{\rm max} = 25 \, h^{-1}$ Mpc. This scale can
be strongly constrained by CMB data -- if the maximal size were
much larger, the voids would add too much additional power (given
the observed value of $F_v$) at the wrong scales
($\ell < 2000$). On the other hand, smaller voids would not be
able to produce any significant contribution to the CMB power spectrum.
The exponent of the size distribution $\alpha$ is weakly constrained
by both theory and observations.  Varying $\alpha$ changes the
position of the peak and the overall power.  This can partially adjust
the influence of the other parameters (see figure \ref{figcl}).

We normalize the distribution by choosing the total number of voids 
so as to fill the required fraction of the universe today, $F_v$.
Redshift surveys point to $F_v \approx 0.4$, ie. 40\% of the volume of
the universe is in underdense regions. The positions of the voids are
then assigned randomly, making sure that they do not overlap. In order
to speed up this process, we consider only a $10^\circ$ cone. This
limits our analysis to $\ell > 100$, which is satisfactory for our
purpose since the main contribution from voids is on much smaller
scales.

\section{Stepping through the void network}
\subsection{Voids between us and the LSS}
\begin{figure}
\centering 
\leavevmode\epsfysize=5cm \epsfbox{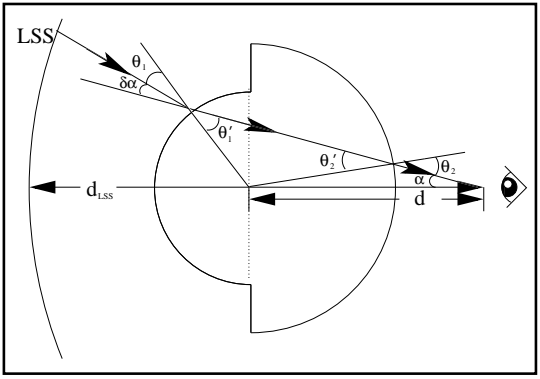}\\ 
\caption[RS]{\label{RS} Cross section of a void (Thompson--Vishniac
model).  The trajectory of a photon is depicted from the LSS to an
observer.  The subscripts 1 and 2 denote quantities at the time the
photon enters the void and at the time it leaves, respectively.
$\alpha$ is defined as the angle formed between the line-of-sight
direction and the direction of the void's centre. $\delta \alpha$ is
defined as the scattering angle of a photon.  $d$ is defined as the
comoving distance of the void's centre. $d_{LSS}$ is defined as the
comoving distance of the LSS.}
\end{figure}

We ray trace photon paths from us to the LSS for the 10$^\circ$ cone
in steps of 1'.  Each void in the present day distribution that is
intersected by the photon path is evolved back in time according to
(\ref{e:vexpand}) to determine whether the photon encounters the
void. If a photon intersects a void between us and the LSS, we compute
the Rees--Sciama (1968) effect due to the deviation in the redshift of
the photon as it passes through the expanding void and the lensing
effect due to the deviation in its path. Thompson \& Vishniac (1987)
applied double local Lorentz transformations at each void boundary to
obtain the redshift deviation (the RS effect) and the scattering angle
of a photon (the lensing effect),
\begin{equation}
\label{e:redshift}
\Delta_{\rm RS} \equiv \left.\frac{\delta T}{T}\right|_{\rm RS} = (H_2R_2)^3 \cos
\theta_2 \left( 3\beta - \frac{2}{3} \cos^2 \theta_2 \right) \,,
\end{equation}
\begin{equation}
\label{e:scatang}
\delta \alpha \equiv -\theta_1 + \theta'_1 +\theta'_2 - \theta_2 = (H_2R_2)^2 \sin(2\theta_2) \,,
\end{equation}
where $H$ is the Hubble parameter, $R \equiv ar_v$ is the proper
length of the void radius, $\beta$ is given by (\ref{e:vexpand}) and
the angles $\theta_1$, $\theta'_1$, $\theta'_2$ and $\theta_2$ are defined by
reference to figure~\ref{RS}.

This treatment agrees with the complementary approach of using the
potential approximation for the under-density
(Mart\'{\i}nez-Gonz\'{a}lez et al. 1990).  The potential approximation
is defined in the EdS background making certain assumptions for which
one of the Einstein equations reduces to the Poisson equation,
\begin{equation}
\label{e:vpois}
\frac{1}{a^2}\nabla^2\psi = 4\pi G\bar\rho \, \delta \,,
\end{equation}
where $\bar \rho$ is the background density and
$\delta=\delta\rho/\bar\rho$ is the density fluctuation field.  For a
spherical void with a thin shell, $\rho(\eta, {\bf x})$ is explicitly
written as,
\begin{equation}
\begin{array}{c}
\rho(\eta, \, r)=\bar\rho(\eta)\theta(r-r_v(\eta))+\rho_{\rm in}(\eta, r)\theta(r_v(\eta)-r)\\+\sigma(\eta)\delta_{\rm Dirac}(r-r_v(\eta)) \,,
\end{array}
\end{equation}
where $\theta$ is the Heaviside function, $\delta_{\rm Dirac}$ is the
Dirac delta function, $\rho_{\rm in}$ is the energy density inside the
void, and $\sigma$ is the surface energy density of the shell. Using
this model of a void and assuming $\rho_{\rm in}$ to be homogeneous,
(\ref{e:vpois}) is easily integrated to give,
\begin{eqnarray}
\label{e:potap}
\psi = \frac{1}{4}H^2a^2\left(r^2-r_v^2\right) \delta \,, \quad
{\rm for} \quad r<r_v \,, \\
\psi =0 \,, \quad {\rm for} \quad r>r_v \,.
\end{eqnarray}
The non-linear growth of the void causes the potential inside it to
grow with respect to the EdS background. This gives rise to the RS
effect given by,
\begin{equation}
\label{e:ISWRS}
\Delta_{\rm RS} = -2\int_{\rm LSS}^0d{\bf x} \cdot\nabla\psi \,.
\end{equation}
The integration of (\ref{e:ISWRS}) for a void between us at the LSS
results in equation (\ref{e:redshift}), that is the Thompson--Vishniac
result.

\subsection{Voids on the LSS}
If a photon intersects a void on the LSS, we use the potential
approximation to calculate the Sachs--Wolfe (1967) effect due to the
photon originating from within the underdensity.  For an empty void
($\delta=-1$) that satisfies the potential approximation
(\ref{e:potap}), the SW effect is given by (Sakai et al. 1999),
\begin{equation}
\label{e:sw}
\begin{array}{c}
\Delta_{\rm SW} \equiv \left.\frac{\delta T}{T}\right|_{\rm SW} = \left.\frac{1}{12}H^2 a^2 \left(r_v^2-r^2 \right)\right|_{\rm LSS} \\ = \frac{1}{12}H^2a^2 \left(r_v^2\cos^2\theta_2-X^2 \right) \,,
\end{array}
\end{equation}
where $X$ is defined as the distance between the centre of the void
and the LSS. Equation (\ref{e:sw}) takes a maximal value at $r=0$
corresponding to the case where a photon originates at the void
centre.

We take into account the finite thickness of the LSS, which suppresses
the SW effect for small voids, by averaging the contribution from a
number of photons originating from a LSS of mean redshift 1100 and
standard deviation in redshift 80.  We also calculate the partial RS
effect (PRS) that arises due to the expansion of the void on the LSS
as the photon leaves it.  Using the potential approximation and
integrating (\ref{e:ISWRS}) we obtain,
\begin{equation}
\begin{array}{c}
\Delta_{\rm PRS} \equiv \left.\frac{\delta T}{T}\right|_{\rm PRS} = 2\left(3 + \frac{\left(r_2^2-r_{\rm LSS}^2\right) + \eta_2\left(\eta_2-4\eta_{\rm LSS}\right)}{\eta^2_{\rm LSS}}\right. \\ \left. + 2\log\left(\frac{\eta_2}{\eta_{\rm LSS}}\right) - 2\frac{r_2 \left(\eta_2 - \eta_{\rm LSS}\right)^2\cos\theta_2 }{\eta_{\rm LSS}^2 \eta_i}\right) \,, 
\end{array}
\end{equation}
where $r_{\rm LSS}$ and $r_2$ are the size of the void at $\eta_{\rm
LSS}$ and at the time the photon leaves the void ($\eta_2$)
respectively and the angle $\theta_2$ is defined by reference to
figure~\ref{RS}.

\begin{figure}
\centering
\leavevmode\epsfysize=6.8cm \epsfbox{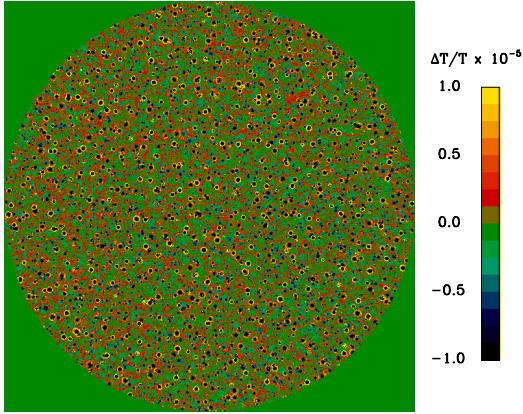}
\caption[fig1]{\label{figmap}The map of the temperature
fluctuations on the surface of last scattering from the
fiducial void model considered. The parameters of this model are
given by $\alpha = 3$, $r_{\rm max}=25 \, h^{-1}$ Mpc and
$F_v=0.4$.}
\end{figure}

\begin{figure}
\centering
\leavevmode\epsfysize=7.4cm \epsfbox{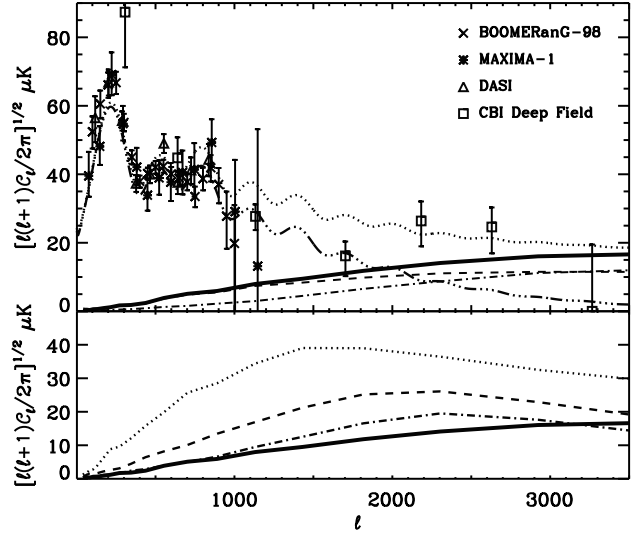}
\caption[fig2]{\label{figcl}Top: The CMB anisotropies produced by the
fiducial void model (solid line) compared to the primary CMB
anisotropies (dashed-triple-dotted). Also plotted are the sum of
primary and void contributions (dotted) as well as the fluctuations
induced purely by voids on the last scattering surface (dashed) and by
those between last scattering and today (dashed-dotted).  We show the
``standard'' cosmological concordance model: of course a combined
analysis of primary and void--induced fluctuations would select a
different cosmology for the primary contribution.

Bottom: Example models depicting a range of void contributions to the 
CMB fluctuations. The models plotted are 
$\alpha = 3$, $r_{\rm max}=25 \, h^{-1}$ Mpc and
$F_v=0.4$ (solid line), $\alpha = 3$, $r_{\rm max}=40 \, h^{-1}$ Mpc and
$F_v=0.4$ (dotted), $\alpha = 3$, $r_{\rm max}=40 \, h^{-1}$ Mpc and
$F_v=0.2$ (dashed) and $\alpha = 6$, $r_{\rm max}=40 \, h^{-1}$ Mpc and
$F_v=0.4$ (dashed-dotted).}
\end{figure}

Once the photon has reached the last scattering surface, we know the
variation of its temperature as well as its position on the LSS and
can create a temperature map (figure \ref{figmap}).  We then use a
flat sky approximation to obtain the $C_\ell$ spectrum of the
anisotropies (White et al. 1999; da Silva 2002) (see figure
\ref{figcl}).  This figure is the main result of this Paper and shows
that a void model motivated by theory and observations can provide
substantial power on scales beyond $\ell = 2000$.

We point out that primordial void parameters are still poorly
constrained by both observation and theory.  The bottom panel of
figure ~\ref{figcl} shows a few further example models.  For a
power-law size distribution (as motivated by the inflationary
scenario), large voids become rarer as $\alpha$ is increased.
Therefore, since void analyses of redshift surveys only sample a
fraction of the volume of the universe, there may exist voids of
larger $r_{\rm max}$ than currently observed.  Models with high
$r_{\rm max}$ tend to predict too much power on scales $\ell \approx
1000$. However, if we take inflationary models with $\alpha > 6$, as
motivated by eg. Occhionero \& Amendola (1994), then the peak moves to
larger $\ell$ and the total power drops. The filling fraction mainly
adjusts the overall power.


\section{Conclusions}
The cosmic microwave background is an excellent tool for probing the
distribution of matter from last scattering until today. In the case
of voids, the strongest signal stems from objects at very high
redshifts, especially from those already present at decoupling.  We
discuss in this Paper the imprint of a power law distribution of
primordial, spherical and compensated voids, which could for example
be generated by a phase transition during inflation.

We show that the signature of such a distribution of voids, that is
compatible with redshift survey observations, contributes additional
power on small angular scales. At the same time, this scenario solves
the void--crisis of the CDM model.  Experiments such as the CBI are
able to directly probe small angular scales and constrain void
parameters. We will present a constraints analysis of a wide range of
void models in a future paper.  In this in--depth analysis, we will
also investigate the non-Gaussian signal of void models that are
compatible with CMB observations as well as any effect of acoustic
waves that primordial voids may propagate to the sound horizon
(Baccigalupi \& Perrotta 1998, Corasaniti et al. 2001).

Other sources are also expected to contribute at high $\ell$.
Probably the strongest of these is the thermal Sunyaev-Zel'dovich (SZ)
effect. Since the thermal SZ effect is strongly frequency--dependent,
experiments which work at about 30 GHz (like CBI) will see a stronger
signal than those working at higher frequencies (Aghanim et
al. 2002). Hence a multi-frequency approach should be able to easily
disentangle the contribution of voids from the SZ
effect. Unfortunately, it seems to be difficult at present to predict
the precise level of the SZ contribution, since different groups are
reporting different results (see eg. Springel et al. 2000 for a
compilation).  Future multi-frequency, high-resolution and high
signal-to-noise maps should be able to significantly constrain the
contribution of primordial voids to the high $\ell$ CMB power
spectrum.  Additionally, deep galaxy redshift surveys and measurements
of the distribution of matter in the Ly-$\alpha$ forest will able to
directly explore the presence of voids in the baryonic matter
distribution at low redshifts.


\section*{Acknowledgments}
It is a pleasure to thank Andrew Liddle for many interesting and
crucial discussions. We also thank Antonio da Silva for helpful
conversations and acknowledge the use of his angular power spectrum
extraction algorithm.  LMG is also grateful to Charles Lineweaver and
the University of New South Wales, where some of this work was carried
out, for their hospitality.  LMG acknowledges support from PPARC.  MK
acknowledges support from the Swiss National Science Foundation.



\begin{thebibliography}{99}
\bibitem{aghanim2002} Aghanim N. {\em et al.}, 2002,
	preprint archived under {\tt astro-ph/0203112}.
\bibitem{aikio98} Aikio J. and M\"ah\"onen P., 1998,
	Astrophys. J. {\bf 497}, 534.
\bibitem{amen96} Amendola L. {\em et al.}, 1996, 
	Phys. Rev. D {\bf 54}, 7199.
\bibitem{amen98} Amendola L., Baccigalupi C. and Occhionero F., 1998,
	Astrophys. J. {\bf 492}, L5.
\bibitem{arnau93} Arnau J.V. {\em et al.}, 1993,
	Astrophys. J. {\bf 402}, 359.
\bibitem{bacci97} Baccigalupi C., Amendola L. and Occhionero F., 1997,
	Mon. Not. R. Astron. Soc. {\bf 288}, 387.
\bibitem{bacci98} Baccigalupi C., 1998,
	Astrophys. J. {\bf 496}, 615.
\bibitem{bacper00} Baccigalupi C., Perrotta F., 2000, 
	Mon. Not. R. Astron. Soc., {\bf 314}, 1
\bibitem{bert85} Bertschinger E., 1985,
	Astrophys. J. Supp. {\bf 58}, 1.
\bibitem{jbd} Brans C. and Dicke C.H., 1961, 
	Phys. Rev. {\bf 24}, 925.
\bibitem{cor01}Corasaniti P.S., Amendola L. and Occhionero F., 2001,
	Mon. Not. R. Astron. Soc., {\bf 323}, 677
\bibitem{dekel86} Dekel A. and Silk J., 1986,
	Astrophys. J. {\bf 303}, 39.
\bibitem{elad96} El-Ad H. {\em et al.}, 1996,
	Astrophys. J. Lett. {\bf 462}, L13.
\bibitem{elad97} El-Ad H., Piran T. and Dacosta L.N., 1997,
	Mon. Not. R. Astron. Soc. {\bf 287}, 790.
\bibitem{elad97b} El-Ad H. and Piran T., 1997,
	Astrophys. J. {\bf 491}, 421.
\bibitem{fullana96} Fullana M.J., Arnau J.V. and Saez D., 1996,
	Mon. Not. R. Astron. Soc. {\bf 280}, 1181.
\bibitem{hoff92} Hoffman Y., Silk J. and Wyse R.F.G., 1992,
	Astrophys. J. Lett. {\bf 388}, L13.
\bibitem{holman90} Holman R., Kolb E.W. and Wang Y., 1990, 
	Phys. Rev. Lett. {\bf 65}, 17.
\bibitem{hoyle2002} Hoyle F. and Vogeley M.S., 2002, 
	Astrophys. J. {\bf 566}, 641.
\bibitem{kauff91} Kauffmann G. and Fairall A.P., 1991,
	Mon. Not. R. Astron. Soc. {\bf 248}, 313.
\bibitem{kauff92} Kauffmann G. and Melott A.L., 1992,
	Astrophys. J. {\bf 393}, 415.
\bibitem{kolb91} Kolb E.W., 1991, in {\em The Birth and Early Evolution
	of the Universe}, Proceedings of the 1990 Nobel Symposium.
\bibitem{la89} La D. and Steinhardt P.J., 1989,
	Phys. Rev. Lett. {\bf 62}, 376.
\bibitem{la89b} La D., Steinhardt P.J. and Bertschinger E., 1989, 
	Phys. Lett. B {\bf 231} 231.
\bibitem{la91} La D., 1991, Phys. Lett. B {\bf 265}, 232.
\bibitem{liddle91}  Liddle A.R. and Wands D., 1991,
	Mon. Not. R. Astron. Soc. {\bf 253}, 637.
\bibitem{liddle92} Liddle A.R. and Wands D., 1992,
	Phys. Lett. B {\bf 276}, 18.
\bibitem{maeda83} Maeda K. and Sato H., 1983,
	Progr. Theor. Phys. {\bf 70}, 772.
\bibitem{mg90a} Mart\'{\i}nez-Gonz\'{a}lez E., Sanz J.L. and Silk J., 1990,
	Astrophys. J. Lett {\bf 355}, 5.
\bibitem{mg90b} Mart\'{\i}nez-Gonz\'{a}lez E. and Sanz J.L., 1990,
	Mon. Not. R. Astron. Soc. {\bf 247}, 473.
\bibitem{cbi02} Mason B.S. et al., 2002, astro-ph/0205384 
\bibitem{mes94} M\'{e}sz\'{a}ros A., 1994,
	Astrophys. J. {\bf 423}, 19.
\bibitem{mes96} M\'{e}sz\'{a}ros A. and Moln\'{a}r Z., 1996,
	Astrophys. J. {\bf 470}, 49.
\bibitem{mull2000} M\"uller V. {\em et al.}, 2000,
	Mon. Not. R. Astron. Soc. {\bf 318}, 280.
\bibitem{occhi94} Occhionero F. and Amendola L., 1994,
	Phys. Rev. D {\bf 50}, 4846.
\bibitem{panek92} Panek M., 1992, Astrophys. J. {\bf 388}, 225.
\bibitem{peacock} Peacock J.A. {\em et al.}, 2001, Nature {\bf 410},
169.
\bibitem{peebles89} Peebles P.J.E., 1989,
	J. R. Astron. Soc. Canada {\bf 83}, 363.
\bibitem{peebles2001} Peebles P.J.E., 2001, 
	Astrophys. J. {\bf 557}, 495.
\bibitem{pelle89} Pellegrini P.S., da Costa L.N. and de Carvalho R.R., 1989,
	Astrophys. J. {\bf 339}, 595.
\bibitem{plio2001} Plionis M. and Basilakos S., 2002,
	Mon. Not. R. Astron. Soc. {\bf 330}, 399.
\bibitem{ryden95} Ryden B.S., 1995,
	Astrophys. J. {\bf 452}, 25.
\bibitem{ryden96} Ryden B.S. and Melott A.L., 1996,
	Astrophys. J. {\bf 470}, 160.
\bibitem{rs1968} Rees M.J. and Sciama D., 1968,
	Nature {\bf 217}, 511 (1968).
\bibitem{sw1967} Sachs R.K. and A.M. Wolfe A.M., 1967,
	Astrophys. J. {\bf 147}, 73.
\bibitem{sakai99} Sakai N., Sugiyama N. and Yokoyama J., 1999,
	Astrophys. J. {\bf 510}, 1.
\bibitem{sato85} Sato H., 1985,
	Prog. Theor. Phys. {\bf 73}, 649.
\bibitem{shi96} Shi X., Widrow L.M. and Dursi L.J., 1996,
	Mon. Not. R. Astron. Soc. {\bf 281}, 565.
\bibitem{silva2002} da Silva A., 2002, DPhil Thesis, University of Sussex 
	(in preparation).
\bibitem{slezak93} Slezak E., de Lapparent V. and Bijaoui A., 1993,
	Astrophys. J. {\bf 409}, 517.
\bibitem{springel00} Springel V., White M. and Hernquist L., 2001,
	Astrophys. J. {\bf 549}, 681; {\em ibid.}
	Astrophys. J. {\bf 562}, 1086.
\bibitem{thomp87} Thompson K.L. and Vishniac E.T., 1987,
	Astrophys. J. {\bf 313}, 517.
\bibitem{turner92} Turner M.S., Weinberg E.J. and Widrow L.M., 1992,
	Phys. Rev. D {\bf 46}, 2384.
\bibitem{white99} White M., Carlstrom J.E., Dragovan M. \&
Holzapfel W.L., 1999, Astrophys. J. {\bf 514}, 12
\bibitem{will01} Will C.M., 2001,
	Living Rev. Rel. {\bf 4}, 4.

\end{thebibliography}
\end{document}